# Chiral quantum magnets with optically and catalytically active spin ladders

Bum Chul Park[1,2,3,4,5]†, Sung-Chul Kim[6]†, Dae Beom Lee[4,5]†, Young Kwang Kim[7], Bomin Kim[8], Sonny H. Rhim[8], Eunsoo Lee[9], Yongju Hong[9], Kwangyeol Lee[9], Sang Hyun Lee[2,3,10], Jessica Ma[1,2,3], Michał Sawczyk[1,2,3], Jun Lu[1,2,3], Jason Manassa[11], Nishkarsh Agarwal,[11] Robert Hovden[11], Sung Ok Won[6], Min Jun Ko[4,5], Minkyu Park[7], Jiung Cho[12], Xiaoming Mao[3,13], Kai Sun[13]*, Young Keun Kim[4,5]*, and Nicholas A. Kotov[1,2,3]*

[1]Department of Chemical Engineering, University of Michigan, Ann Arbor, MI, USA.

[2]Biointerfaces Institute, University of Michigan, Ann Arbor, MI, USA.

[3] Center for Complex Particle Systems (COMPASS), University of Michigan, Ann Arbor, MI, USA.

[4]Department of Materials Science and Engineering, Korea University, Seoul, KR.

[5]Brain Korea Center for Smart Materials and Devices, Korea University, Seoul, KR

[6] Advanced Analysis and Data Center, Korea Institute of Science and Technology, Seoul, KR.

[7]Virtual Lab Inc., Seoul, KR.

[8]Department of Physics, University of Ulsan, Ulsan, KR.

[9]Department of Chemistry and Research Institute for Natural Science, Korea University, Seoul, KR.

[10]Department of Electrical Engineering and Computer Science, University of Michigan, Ann Arbor, MI, USA.

[11]Department of Materials Science and Engineering, University of Michigan, Ann Arbor, MI, USA

[12] Department of Materials Science and Engineering, Hongik University, Sejong, KR.

[13]Department of Physics, University of Michigan, Ann Arbor, MI, USA

*Corresponding authors. E-mail addresses: sunkai@umich.edu (K. Sun), kim97@korea.ac.kr (Y. K. Kim), kotov@umich.edu (N. A. Kotov)

†These authors made equal contributions to this study.

**ABSTRACT:**

Chiral quantum magnets with spin-states separated by large energy gap are technologically attractive but difficult to realize. Geometrically frustrated topological states with nanoscale chirality may offer a chemical pathway to such materials. However, room temperature spin misalignment, weakness of Dzyaloshinskii–Moriya interactions, and high energy requirements for lattice distortions set high physicochemical barriers for their realization. Here, we show that layered iron oxyhydroxides (LIOX) address these challenges due to chirality transfer from surface ligands into spin-states of dimerized $FeO_6$ octahedra with zig-zag stacking. The intercalation of chiral amino acids induces angular displacements in the antiferromagnetic spin pairs with a helical coupling of magnetic moments along the screw axis of the zig-zag chains, or helical spin-ladders. Unlike other chiral magnets, the spin states in LIOX are chemically and optically accessible: they display strong optical resonances with helicity-matching photons and enable spin-selective charge transport. The static rather than dynamic polarization of spin ladders in LIOX makes them particularly suitable for catalysis. Room-temperature spin pairing, field-tunability, environmental robustness, and synthetic simplicity make LIOX and its intercalates uniquely practical family of quantum magnets.

**INTRODUCTION:**

Studies of quantum effects in low-dimensional nanostructures [1–3] have largely focused on the emission of entangled photons from quantum-confined optical materials [4–6]. In contrast, quantum phenomena in magnetic materials, such as topological spin phases [7–11], are more challenging to observe, as discrete magnetic spin states typically have much smaller energy gaps than their electronic counterparts. According to the Mermin-Wagner theorem, low-dimensional spin systems should display enlarged energy gaps between spin states due to topological constraints on spin coupling, an effect often referred to as the Haldane gap [7,8]. The experimental observation of magnetic topological states often requires, however, extreme conditions, including near absolute zero temperatures [11], high magnetic fields [12] specialized imaging tools [13,14], and near-perfect crystallinity [15]. Although room-temperature spin states are possible for skyrmions [16–18], the requirement to sandwich them between layers metal opaque in the visible range significantly limits both optical interrogation and chemical accessibility. Although technologically and fundamentally attractive [19], room temperature quantum magnets that are simultaneously optically active and chemically accessible remain unknown.

Chiral nanowires, nanoribbons, nanoplatelets, and nanosheets [20–22] combine mirror asymmetry with strong spatial confinement, enabling a widened Haldane gap suitable for room-temperature quantum magnets. These and related chiral nanostructures can form stable dispersions and exhibit spin-dependent chiroptical transitions in the visible range [23–25] and spin-selective charge transport, known as chirality-induced spin selectivity effect (CISS) [26]. The realization of chirality-enabled quantum magnets should also be feasible in layered nanomaterials with large basal spacing, where spins are coupled along curved planes or helical chains, while remaining decoupled from adjacent atoms. Notably, both types of chiral nanostructures are expected to support preferential spin alignment in their ground state without the high magnetic fields that are typical for many studies of magnetic quantum effects. Direct chemical access to these helical states can be synergistically combined with CISS effect; they can also be integrated into spintronic junctions [27] to enhance overall spin polarization.

Prior studies of transition metal oxides [28,29] and nanoscale assemblies [30], clearly indicate that the synthesis of chiral quantum magnets with optical activity and (electro)chemically accessible interfaces is non-trivial. On one hand, the rigidity of metallic and ceramic crystal lattices creates high energy barriers for converting achiral structures into chiral ones [31]. On the other hand, materials with more deformable crystal lattices, including metal oxides, sulphides, and hybrid perovskites, are typically antiferromagnetic and, according to the Rosenfeld equation [32], they will have near-zero chiroptical activity. Guided by these theoretical and experimental findings, we chose iron oxyhydroxides (FeOOH) to explore the emergence of the Haldane gap. Interconvertible FeOOH polymorphs [33,34] are known to favour the formation of nanosheets and nanoribbons [33]. We hypothesized that the orientation of and spin coupling between interconnected octahedra producing FeOOH compounds can be modulated by chiral surface ligands that could induce quantum hybridization with the electronic states of magnetic atoms and cause chirality-induced bond distortion. Additional advantages of using iron-based materials are their Earth abundance compared to rare-earth elements [35].

**Molecular design of the chiral low-dimensional spin system**

Layered iron hydroxide (LIOX) represents a new phase of FeOOH that emerges when the magnetite ($Fe_3O_4$) phase becomes thermodynamically unstable (**Figs. 1a, b**; Supplementary Figs. 1–3; and Supplementary Text 1) [36]. Its formation proceeds via oriented attachment (**Fig. 1c, d**) [37], a process highly dependent on surface ligand chemistry (Supplementary Fig. 4, 5; and Supplementary Text 1). Departing from the commonly used thiols in chiral nanostructure synthesis [23,38], we selected carboxyl containing surface ligands, such as acetate and amino acids, because they offer two key advantages: (1) the ability to form bidentate coordination bonds with $Fe^{3+}$ and $Fe^{2+}$ ions and (2) facile incorporation into the layered FeOOH crystal lattice. This two-point binding induces greater lattice distortion than single-point ligands. It allows for variable hybridization between organic and inorganic electronic states, providing diverse synthetic routes to engineer quantum spin states. Furthermore, these ligands are interchangeable, enabling the development of a family of chiral nanomaterials with compositional tunability analogous to hybrid perovskites (Supplementary Fig.1).

After 8 hours of reaction, we obtained disc-shaped nanoplatelets with an average diameter of 26.1 ± 3.3 nm and a thickness of 18.3 ± 2.7 nm (**Fig. 2a, b**; and Supplementary Fig. 5). The nanoplatelets exhibited a bright red-orange colour and showed distinct layering, with a large basal spacing of 1.14 nm (**Fig. 2c, d**). This interlayer separation exceeds the critical interatomic distance required for magnetic exchange interactions[39], thereby reducing interplane coupling between iron atoms, an essential feature for spin-state confinement. It does not diminish with the acetate exchange to other ligands and can be expanded even further (Supplementary Figs. 5 f–h; and Supplementary Movie 1). The synthesis can be readily scaled to 100 g batches (Supplementary Fig. 2).

Cryo-TEM microscopy, 3D reconstructions of the LIOX particles (Supplementary Fig. 5; Supplementary Movie 1), high-resolution X-ray powder diffraction (HRXRPD, **Fig. 2e**) confirms the single-phase nature of the synthesized nanostructures (Supplementary Fig. 6). The crystal lattice of LIOX differs from all known polymorphs of FeOOH and other iron oxides (Supplementary Fig. 7). Raman scattering spectra from the nanoplatelets reveal strong Fe–O lattice/phonon vibrations that are distinct from those of α–, β–, γ–, and δ–FeOOH (**Fig. 2f**; and Supplementary Fig. 8) [40]. Sharp peaks at 130, 175, 234, 328, 409, 459, and 650 cm$^{-1}$ indicate a well-defined solitary site symmetry of the Fe–O bonds, consistent with FeO$_6$ octahedra and Fe–acetate coordination (Supplementary Table 1). The Fourier transform infrared (FT–IR) spectra reveal $\Delta(v_{as}(COOFe)-v_s(COOFe))$ peak separation of 134 cm$^{-1}$, which indicates bidentate bridging between acetate and Fe atoms and multipoint attachment (**Fig. 2g**; and Supplementary Table 2). The Fe K–edge of X-ray absorption near–edge structure (XANES) spectrum supports the presence of the ferric ion (Fe$^{3+}$) state (**Fig. 2h**). The cumulative analysis of X-ray photoelectron spectroscopy (XPS, Supplementary Fig. 9; and Supplementary Table 3), Rutherford backscattering spectrometry/time of flight–elastic recoil detection (RBS/TOF–ERD, Supplementary Fig. 10), thermogravimetric analysis/differential thermal analysis (TGA/DTA, Supplementary Fig. 11), and in situ thermal XRPD (Supplementary Fig. 12) indicate that LIOX is a new polymorph of FeOOH where CH$_3$COO$^-$ ions and FeOOH units are combined in a 2:1 molar ratio (Supplementary Fig. 13).

*Ab initio* computational analysis of HR-XRPD diffractograms allowed us to establish the crystallographic features and atomic coordinates of LIOX (**Fig. 2e, i**; Supplementary Tables 4, 5; and Supplementary Text 2). This polymorph of FeOOH belongs to a monoclinic *Pn* group with lattice parameters $a$ = 3.85 Å, $b$ = 11.49 Å, $c$ = 9.82 Å, and $β$ = 88.47 Å, where $b$ is the basal spacing along [010] axis. The diffraction pattern of this previously unreported FeOOH polymorph shows an intense (*0k0*) reflection characteristic of planar crystallites. Notably, $FeO_6$ octahedra in LIOX are dimerized by bidentate bridging from the surface ligands, essential for their control of spin coupling. When viewed from the [100] direction, the zig-zag arrangement of dimerized $FeO_6$ with *n*-glide symmetry becomes apparent. $FeO_6$ octahedra share an edge with the neighbouring dimer and form the *ac* plane of the nanosheets (Supplementary Fig. 14). These structural features indicate topological frustrations and differentiate LIOX from γ and 'green rust' forms of FeOOH with simple zig-zag or planar structural motifs (Supplementary Figs. S15–18).

We also assessed the structural stability of LIOX by DFT calculations with generalized gradient approximation Perdew–Burke–Ernzerhof (GGA–PBE) pseudopotentials and *ab initio* molecular dynamics (AIMD) (Supplementary Figs. 19–21; Supplementary Table 6; and Supplementary Movie 2). The structural framework of the LIOX was maintained without breakage of the $FeO_6$ arrangement; at temperatures as high as 300 K, no changes in basal spacing were observed.

**Ligand-induced chirality of the octahedral layers**

*L*– and *D*–enantiomers of proline (Pro) can replace acetate in LIOX via simple one-step process of ion exchange. The ionized enantiomers of the amino acid diffuse through the wide basal spacing between layers of $FeO_6$ octahedra penetrating deep inside the nanoparticles (**Fig. 3a**; Supplementary Figs. 5f–h, 22; and Supplementary Movie 1). The chemical bonding of Pro to the octahedral pairs renders this material chiral via twist-distortion of the octahedral lattice and hybridization of quantum states of the inorganic lattice with those of organic ligands. Vibrational circular dichroism (VCD), together with infrared absorption spectra, reveals a distinct mirror-symmetric band at 1583 cm$^{-1}$, corresponding to the asymmetric vibration $v_{as}$(COO) of Pro-Fe coordination (**Fig. 3b**). The VCD activity at this frequency originates from the twist of the crystal lattice induced by the bidentate binding of Pro enantiomers. HR-XRPD data from *L*-Pro LIOX further confirm this structural distortion. Specifically, Pro coordination disrupts the mirror symmetry of the *n*-glide *ac* plane in the LIOX sublattice, introducing $2_1$ screw axes aligned along the zig-zag topography (**Fig. 3a**; Supplementary Fig. 23; and Supplementary Tables 7, 8).

MD and DFT simulations provided insights into how crystallographic asymmetry is influenced by the binding of chiral ligands (**Fig. 3c**). Pro ligands induce distinct helical distortions along the zig-zag chains of FeO$_6$ octahedra. The low-dimensionality of the octahedral layers allows this distortion to propagate within the unit cells without encountering the high energy barrier characteristic of 3D crystals and other 2D materials [23,41].

To quantify chiral symmetry breaking in the Fe-O octahedral framework, we used the Hausdorff chirality measure (HCM, Supplementary Fig. 24)[42]. HCM increases in LIOX phases modified with either enantiomer of Pro, gradually increasing during structural relaxation. This finding indicates the progressive twist-deformation and related chirality increase of the zig-zag chains (**Fig. 3d**).

**Quantum magnetism of LIOX**

As extensively demonstrated in previous studies [13,14,18,43,44], lattice chirality induces Dzyaloshinskii–Moriya interactions (DMI), giving rise to magnetic phases with a quantum spin gap. For LIOX, a screw axis symmetry of the atomic lattices (**Fig. 3a, c**) is expected to induce helicity in neighbouring spin dimers, rendering the magnetic state of the entire material chiral, such that $\langle \vec{n} \cdot \vec{S}_1 \times \vec{S}_2 \rangle \neq 0$. Furthermore, the increased basal spacing in the crystal lattice enhances spin state quantization (**Fig. 4a**)[10,11] by decreasing inter-sheet spin coupling. Together, these structural features—lattice chirality, screw-axis symmetry, and expanded interlayer distance— synergistically stabilize topological spin-states in LIOX.

DFT calculations were carried out to understand better the nature of magnetic states in achiral and chiral forms of LIOX (**Fig. 4b, c**; Supplementary Fig. 25; and Supplementary Table 9). Unlike 3D crystals exhibiting long-range Néel order (i.e., 3D Heisenberg antiferromagnets), achiral and chiral LIOX display low-dimensional antiferromagnetism. As hypothesized, the inter-sheet spin coupling is suppressed by approximately three orders of magnitude compared to in-plane spin coupling. The latter is defined by three constants describing spin coupling between corner ($J_1$), edge ($J_2$), and face-sharing ($J_3$) octahedra, all of which are negative, indicating antiferromagnetic antiparallel spin alignment, driven by super-exchange interactions through $Fe^{3+}$–O–$Fe^{3+}$ linkages.

Notably, the configuration and organization of spin-states differ strongly between achiral and chiral forms of LIOX while the size and geometry of nanoparticles do not. The spins in achiral LIOX, exhibits a semi-dimensional ladder structure where $J_3 > J_2$, with 'legs' and 'rungs' formed by spin pairs coupled through $J_1$ and $J_2$, respectively (**Fig. 4b**). Conversely, chiral LIOX adopts a mirror-asymmetric spin-state where $J_1 > J_3$, resulting in a staggered spin ladder defined by $J_2$ and $J_1$ interactions (**Fig. 4c**). These spin configurations distinguish LIOX from nanoscale or microscale 3D metal oxides, antiferromagnets, magnetic multilayers, spin liquids, and isolated skyrmions. The magnetic susceptibility measurements of LIOX reveal an increase with temperature above a magnetic transition at 75 K, reaching a broad maximum near 400 K (**Fig. 4d, e**; and Supplementary Fig. 26). This behavior is consistent with the theoretical model of a quantum Haldane spin-gap ($\Delta$). The latter can be calculated as

$$\chi_{spin}(T) = aT^{-1/2}\exp(-\Delta E/k\,T) \qquad (1)$$

where $\Delta E$, $k$, and $T$ are the spin-gap, Boltzmann constant, and temperature, respectively; $a$ is the proportionality constant related to excitation energy dispersion[11,45]. Applying this formalism, the spin-gaps for acetate LIOX, *L*-Pro LIOX, and *D*-Pro LIOX are estimated to be approximately 30.4, 36.3, and 37.0 meV, respectively (Supplementary Fig. 27), indicating a widening of the Haldane gap due to chirality of the nanostructures. These findings are corroborated by DFT calculations showing enhanced spin coupling in chiral LIOX (Supplementary Table 9). The ground state corresponds to a spin singlet ($S = 0$) with the configuration of $(|\uparrow\downarrow\rangle - |\downarrow\uparrow\rangle)/\sqrt{2}$, while the excited states are spin triplets ($S = 1$) with configurations of $|\uparrow\uparrow\rangle, |\downarrow\downarrow\rangle$, and $(|\uparrow\downarrow\rangle + |\downarrow\uparrow\rangle)/\sqrt{2}$ (**Fig. 4f**). The net magnetization of these states may manifest in both chemical and optical properties, which could be useful for further experimental validation and practical utilization of these quantum spin-states.

**Optical and electrocatalytic properties of quantum spin ladders**

Colloidal dispersions of LIOX functionalized with acetate, *L*–Pro, and *D*–Pro exhibit mirror-image magnetic circular dichroism (MCD) spectra in the ultraviolet (UV) and visible ranges, indicating strong ligand-lattice hybridization of electronic states (**Fig. 5a, b**; and Supplementary Figs. 28, 29)[46,47]. Pro enantiomers remove the degeneracy between left- and right-handed spin-states, due to the twisted zig-zag arrangement of FeO$_6$ octahedra. Remarkably, even when the Pro ligands are exchanged with achiral 4-(phenylazo)benzoic acid (PABA), a strong CD response persists, suggesting that the chiroptical activity originates from the helicity of the zig-zag FeO$_6$ chains, initially induced by the chirality transfer from Pro (Supplementary Figs. 30, 31). Identical chiroptical effects were observed for both *L*– and *D*–Asp (Supplementary Fig. 32), indicating that this phenomenon is generalizable across multiple chiral ligands. Note that the positive surface charge of LIOX nanoparticles ($\zeta$ = +39 mV) favors the intercalation and exchange of neutral or negatively charged enantiomers (Supplementary Fig. 33). As a control, lepidocrocite—a non-layered iron oxide with no basal spacing—showed no CD response after treatment with Pro (Supplementary Fig. 34), highlighting the importance of basal spaces chirality transfer and as access routes the spin-states in LIOX.

MCD spectra of LIOX functionalized with *L*– and *D*–Pro show a 30% enhancement under an applied magnetic field in the N–S direction and a corresponding 30% decrease when the field is reversed (S–N direction, **Fig. 5b**). This response arises from thermally accessible triplet excitations of the topological spin ladders. These excitations increase net magnetization under an external magnetic field, providing an additional mechanism for tuning the chiroptical response. The degree of polarization (*DP*) under a magnetic field result from the combined effects of Zeeman splitting and the intrinsic optical activity of chiral LIOX, which can be calculated as:

$$DP = \frac{g_{eff}\mu_B B}{kT} + DP_0 \qquad (2),$$

where $g_{eff}$ is the effective gyromagnetic ratio, $\mu_B$ is the Bohr magneton, *B* is the applied magnetic field, $DP_0$ is the degree of spin-polarized absorption of chiral LIOX in the absence of the magnetic field [25]. The contribution of the Zeeman splitting in chiral LIOX can be quantified using the effective gyromagnetic ratio ($g_{eff}$), which is 19, 21, and 23 for acetate, *L*-Pro LIOX, and *D*-Pro LIOXs, respectively (**Fig. 5c**; Supplementary Fig. 35; Supplementary Table 10; and Supplementary Text 3, 4). These values are comparable to the giant Zeeman splitting observed in semiconductors with ferromagnetic coupling.

The increased magnetic susceptibility of LIOX enables stronger Zeeman effects in the excitation of spin-gaps, allowing for magnetic field-tuneable selectivity toward photons with opposite spin angular momentum (i.e., circular polarization). This field-induced optical activity is particularly notable given that the discrete energy state of the quantum antiferromagnetic spin gap is observed at room temperature and in aqueous dispersions. For comparison, goethite, a 3D antiferromagnet without discrete energy levels, displays negligible MCD under similar conditions (Supplementary Fig. 36).

Because the ground state energies of the left- and right-handed spin ladders in chiral nanostructures differ, electrons exchanged at their interfaces will be spin-polarized. This spin polarization arises from chiral magnetic states and is additive to the CISS effect [26], which occurs when electrons pass through molecules with chiral electronic states. The chemically accessible interface of LIOX particles provide a unique opportunity to evaluate the utility of room-temperature Haldane gaps in electrocatalytic oxygen evolution reactions (OER), which are known to be sensitive to spin polarization of electrons due to triplet oxygen forming as an intermediate (Supplementary Text 5; and Supplementary Figs. 37–39).

LIOX's large basal spacing also facilitates direct water interaction with the spin ladders, enhancing catalytic accessibility. Compared to conventional 3D FeOOH polymorphs, LIOX shows superior OER performance (**Fig. 5d**; and Supplementary Fig. 37). Notably, chiral LIOX materials exhibit significantly lower Tafel slopes than their achiral counterparts, indicating enhanced catalytic kinetics (electron transfer): 40.3 mV dec$^{-1}$ for acetate LIOX, 21.8 mV dec$^{-1}$ for *L*-Pro LIOX, and just 18.9 mV dec$^{-1}$ for *D*-Pro LIOX (**Fig. 5e**; and Supplementary Fig. 38). External magnetic field further reduces the Tafel slope of achiral LIOX by ~30%, even at a weak magnetic strength of 0.2 T (**Fig. 5f**; and Supplementary Fig. 39). Because spin ladders represent the ground state of the system, they enable stronger spin polarization than typical CISS systems—even at much higher magnetic field strengths (Supplementary Table 11) [48–50].

In conclusion, the chirality of surface ligands transforms the zig-zag chains of achiral, low-dimensional LIOX into room-temperature quantum magnets. The dispersibility and chemical accessibility of octahedral nanosheets comprising LIOX enable observation of strong chiroptical activity in isotropic dispersion and OER, leveraging spin-polarized ground states configured as helical spin-ladders. The environmental robustness, earth-abundance of iron, and synthetic simplicity makes LIOX and its derivatives a uniquely practical family of chiral quantum magnets. Based on antiferromagnetically coupled spin pairs, its structural design offers a versatile chemical platform that can be extended to other low-dimensional magnetic materials. The multiplicity of amino acids available for LIOX intercalation also open promising avenues for further studies for asymmetric catalysis.

**References**

1. Li, J., Li, X. & Zhu, H. Symmetry engineering in low-dimensional materials. *Mater. Today* **75**, 187–209 (2024).

2. Chen, Y. *et al.* Multidimensional nanoscopic chiroptics. *Nat. Rev. Phys.* **4**, 113–124 (2022).

3. Ma, W. *et al.* Chiral Inorganic Nanostructures. *Chem. Rev.* **117**, (2017).

4. García De Arquer, F. P. *et al.* Semiconductor quantum dots: Technological progress and future challenges. *Science* **373**, eaaz8541 (2021).

5. Aharonovich, I., Englund, D. & Toth, M. Solid-state single-photon emitters. *Nat. Photonics* **10**, 631–641 (2016).

6. He, Y.-M. *et al.* Single quantum emitters in monolayer semiconductors. *Nat. Nanotechnol.* **10**, 497–502 (2015).

7. Haldane, F. D. M. CONTINUUM DYNAMICS OF THE 1-D HEISENBERG ANTIFERROMAGNET: IDENTIFICATION WITH THE 0(3) NONLINEAR SIGMA MODEL. *Phys. Lett.* **93**, (1983).

8. Haldane, F. D. M. Nonlinear Field Theory of Large-Spin Heisenberg Antiferromagnets: Semiclassically Quantized Solitons of the One-Dimensional Easy-Axis Néel State. *Phys.*

*Rev. Lett.* **50**, 1153–1156 (1983).

9. Broholm, C. *et al.* Quantum spin liquids. *Science* **367**, eaay0668 (2020).

10. Lake, B. *et al.* Confinement of fractional quantum number particles in a condensed-matter system. *Nat. Phys.* **6**, 50–55 (2010).

11. Dagotto, E. & Rice, T. M. Surprises on the Way from One- to Two-Dimensional Quantum Magnets: The Ladder Materials. *Science* **271**, 618–623 (1996).

12. Novoselov, K. S. *et al.* Room-Temperature Quantum Hall Effect in Graphene. *Science* **315**, 1379–1379 (2007).

13. Moreau-Luchaire, C. *et al.* Additive interfacial chiral interaction in multilayers for stabilization of small individual skyrmions at room temperature. *Nat. Nanotechnol.* **11**, 444–448 (2016).

14. Camley, R. E. & Livesey, K. L. Consequences of the Dzyaloshinskii-Moriya interaction. *Surf. Sci. Rep.* **78**, 100605 (2023).

15. Yang, J. *et al.* Robust Two-Dimensional Ferromagnetism in $Cr_5Te_8$/$CrTe_2$ Heterostructure with Curie Temperature above 400 K. *ACS Nano* **17**, 23160–23168 (2023).

16. Zhang, H. *et al.* Room-temperature skyrmion lattice in a layered magnet (Fe0.5Co0.5)5GeTe2. *Sci. Adv.* **8**, eabm7103.

17. Zhang, C. *et al.* Above-room-temperature chiral skyrmion lattice and Dzyaloshinskii–Moriya interaction in a van der Waals ferromagnet Fe3−xGaTe2. *Nat. Commun.* **15**, 4472 (2024).

18. Chen, G. & Schmid, A. K. Imaging and Tailoring the Chirality of Domain Walls in Magnetic Films. *Adv. Mater.* **27**, 5738–5743 (2015).

19. Aiello, C. D. *et al.* A Chirality-Based Quantum Leap. *ACS Nano* **16**, 4989–5035 (2022).

20. Amabilino, D. B. *Chirality at the Nanoscale: Nanoparticles, Surfaces, Materials and More*. (2009).

21. Bisoyi, H. K. & Li, Q. Light-directing chiral liquid crystal nanostructures: from 1D to 3D. *Acc. Chem. Res.* **47**, 3184–95 (2014).

22. Gao, X., Han, B., Yang, X. & Tang, Z. Perspective of chiral colloidal semiconductor nanocrystals: opportunity and challenge. *J. Am. Chem. Soc.* **141**, 13700–13707 (2019).

23. Yeom, J. *et al.* Chiromagnetic nanoparticles and gels. *Science* **359**, 309–314 (2018).

24. Qian, Q. *et al.* Chiral molecular intercalation superlattices. *Nature* **606**, 902–908 (2022).

25. Long, G. *et al.* Spin control in reduced-dimensional chiral perovskites. *Nat. Photonics* **12**, 528–533 (2018).

26. Naaman, R., Paltiel, Y. & Waldeck, D. H. Chiral molecules and the electron spin. *Nat. Rev. Chem.* **3**, 250–260 (2019).

27. Ohno, Y. *et al.* Electrical spin injection in a ferromagnetic semiconductor heterostructure. *Nature* **402**, 790–792 (1999).

28. Kumar, P. *et al.* Photonically active bowtie nanoassemblies with chirality continuum. *Nature* **615**, 418–424 (2023).

29. Li, C. *et al.* Ultrasmall Magneto-chiral Cobalt Hydroxide Nanoparticles Enable Dynamic Detection of Reactive Oxygen Species *in Vivo*. *J. Am. Chem. Soc.* **144**, 1580–1588 (2022).

30. Li, Z. *et al.* A magnetic assembly approach to chiral superstructures.

31. Hananel, U., Ben-Moshe, A., Diamant, H. & Markovich, G. Spontaneous and directed symmetry breaking in the formation of chiral nanocrystals. *Proc. Natl. Acad. Sci.* **116**, 11159–11164 (2019).

32. Rosenfeld, L. Quantenmechanische Theorie der natürlichen optischen Aktivität von Flüssigkeiten und Gasen. *Z. Für Phys.* **52**, 161–174 (1929).

33. Puthirath Balan, A. *et al.* Exfoliation of a non-van der Waals material from iron ore hematite. *Nat. Nanotechnol.* **13**, 602–609 (2018).

34. Park, B. C. *et al.* Surface-ligand-induced crystallographic disorder–order transition in

oriented attachment for the tuneable assembly of mesocrystals. *Nat. Commun.* **13**, 1144 (2022).

35. Vasiliev, A., Volkova, O., Zvereva, E. & Markina, M. Milestones of low-D quantum magnetism. *Npj Quantum Mater.* **3**, 18 (2018).

36. Park, B. C. *et al.* Bioinspired redox-coupled conversion reaction in FeOOH-acetate hybrid nanoplatelets for Na ion battery. *J. Mater. Chem. A* **10**, 17740–17751 (2022).

37. Tang, Z., Kotov, N. A. & Giersig, M. Spontaneous Organization of Single CdTe Nanoparticles into Luminescent Nanowires. *Science* **297**, 237–240 (2002).

38. Liu, X. *et al.* Histidine-Mediated Synthesis of Chiral Cobalt Oxide Nanoparticles for Enantiomeric Discrimination and Quantification. *Small* **19**, (2023).

39. Antoniak, C. *et al.* Composition dependence of exchange stiffness in Fe x Pt 1 − x alloys. *Phys. Rev. B* **82**, 064403 (2010).

40. De Faria, D. L. A., Venâncio Silva, S. & De Oliveira, M. T. Raman microspectroscopy of some iron oxides and oxyhydroxides. *J. Raman Spectrosc.* **28**, 873–878 (1997).

41. Lu, J. *et al.* Nano-achiral complex composites for extreme polarization optics. *Nature* **630**, 860–865 (2024).

42. Buda, A. B. & Mislow, K. A Hausdorff chirality measure. *J. Am. Chem. Soc.* **114**, 6006–6012 (1992).

43. Chen, G. *et al.* Observation of Hydrogen-Induced Dzyaloshinskii-Moriya Interaction and Reversible Switching of Magnetic Chirality. *Phys. Rev. X* **11**, 021015 (2021).

44. Yang, H., Boulle, O., Cros, V., Fert, A. & Chshiev, M. Controlling Dzyaloshinskii-Moriya Interaction via Chirality Dependent Atomic-Layer Stacking, Insulator Capping and Electric Field. *Sci. Rep.* **8**, 12356 (2018).

45. Troyer, M., Tsunetsugu, H. & Würtz, D. Thermodynamics and spin gap of the Heisenberg ladder calculated by the look-ahead Lanczos algorithm. *Phys. Rev. B* **50**, 13515–13527

(1994).

46. He, Y. P. *et al.* Size and structure effect on optical transitions of iron oxide nanocrystals. *Phys. Rev. B* **71**, 125411 (2005).

47. Li, C. *et al.* Chiral Iron Oxide Supraparticles Enable Enantiomer-Dependent Tumor-Targeted Magnetic Resonance Imaging. *Adv. Mater.* **35**, (2023).

48. Vadakkayil, A. *et al.* Chiral electrocatalysts eclipse water splitting metrics through spin control. *Nat. Commun.* **14**, 1067 (2023).

49. Garcés-Pineda, F. A., Blasco-Ahicart, M., Nieto-Castro, D., López, N. & Galán-Mascarós, J. R. Direct magnetic enhancement of electrocatalytic water oxidation in alkaline media. *Nat. Energy* **4**, 519–525 (2019).

50. Ren, X. *et al.* Spin-polarized oxygen evolution reaction under magnetic field. *Nat. Commun.* **12**, 2608 (2021).

**Data availability:** The authors declare that the data supporting the findings of this study are available in the article and its supplementary information files. Source data are provided with this paper.

**Acknowledgements:** This work was supported by the National Research Foundation of Korea funded by the Ministry of Science and ICT (RS-2024-00347718) to Y.K.K, and the Office of Naval Research (MURI N00014-20-1-2479) to N.A.K, K.S. and X.M. and the Office of Naval Research (N00014-21-1-2770), and the Gordon and Betty Moore Foundation (N031710) to K.S. and NSF 2243104, and the center of Complex Particle Systems (COMPASS) to N.A.K. and X.M. R.H. acknowledges support from the U.S. Department of Energy, Basic Energy Sciences, under award DE-SC0024147.

**Author contributions:** B.C.P, S.K. and D.B.L. contributed equally to this study and wrote the original draft. B.C.P., Y.K.K., and N.A.K. envisaged and designed the experiments. S.K., B.C.P., and Young Kwang K. solved the unknown structural model. B.C.P., D.B.L., M.J.K., and T.M.K. synthesized the LIOX. S.K., S.O.W., and J.C. ran the X-ray analysis. J.M. and R.H. observed electron microscopy. Young Kwang K., M.P., J.M., and M.S. ran the computational calculation. B.C.P., K.S., X.M., B.K., and S.H.R. analyzed the quantum magnetism. B.C.P., S.H.L., J.L., and J.M. examined the chiroptics. E.L, Y.H., and K.L. performed the electrochemical catalyst analysis. N.A.K., Y.K.K., and K.S. conceived and supervised the project.

**Competing interests:** The authors declare no competing interests.

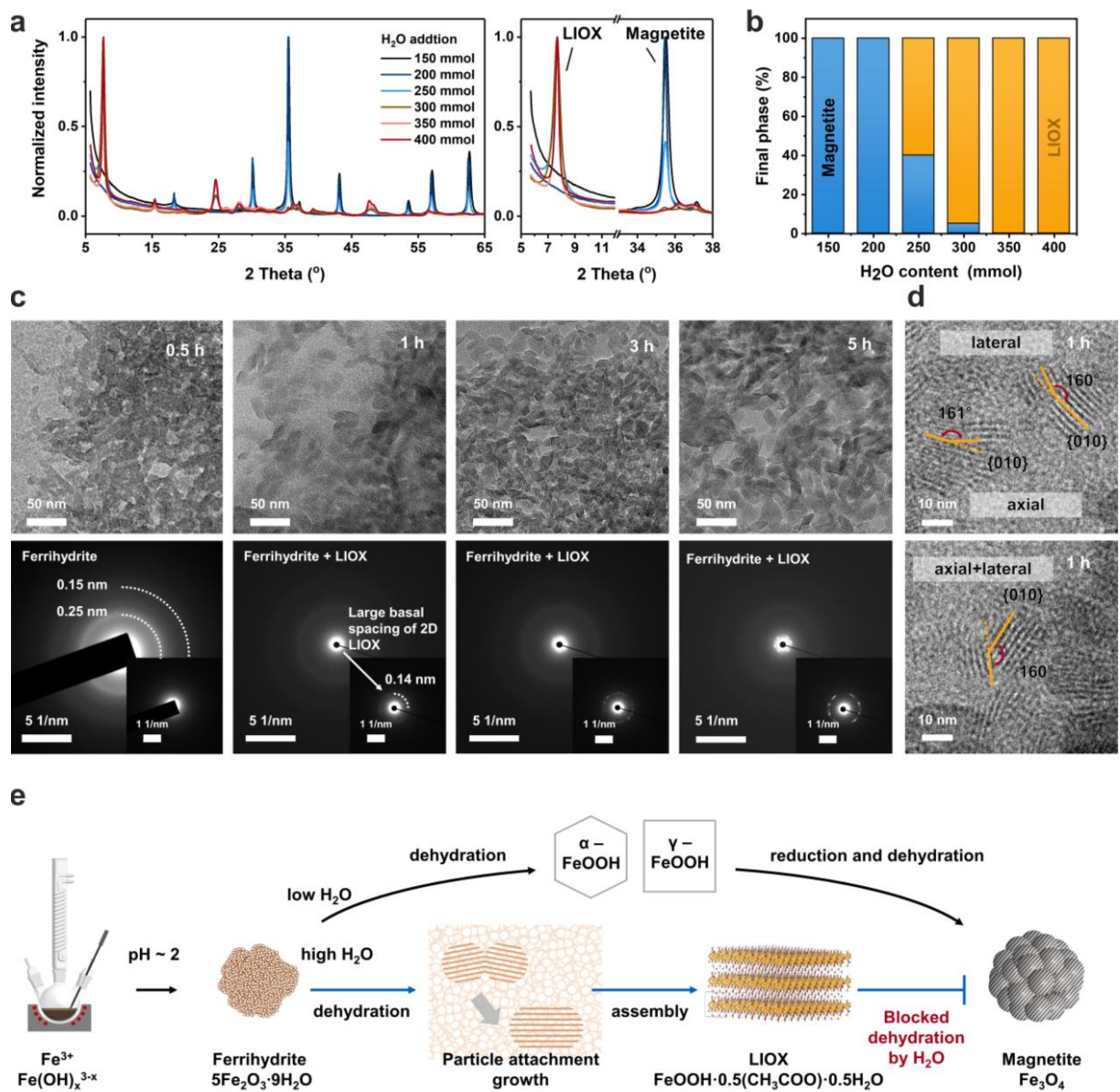

**Fig. 1. Crystallization process and two-dimensional nature of LIOX nanoplatelets obtained by self-assembly of ferrihydrite nanoparticles.** (a) XRPD of the samples synthesized with different H₂O contents from 150 to 400 mmol. As H₂O increases, a sharp diffraction peak near 2θ = 7° emerges, which is attributed to 2D FeOOH with a large basal spacing. (b) The ratio between 2D FeOOH and magnetite calculated by Rietveld's quantitative analysis. For crystallographic data of magnetite, PDF No, 04–002–0618 was used. (c) EX-situ TEM images and SAED patterns show the stepwise transformation of LIOX during 5 h of reaction. The value of d-spacings measured in SAED patterns indicate the formation of ferrihydrite intermediate (t = 0.5 h) and LIOX (from t = 1 h). (d) Magnified TEM images to observe particle attachment growth at the early stage of reaction time (t = 0.5). The orange line and red lines highlight the alignment event between nanoplatelets. (e) Proposed crystallization pathway for 2D FeOOH nanostructures. With low H₂O volume (black lines), ferrihydrite transformed into magnetite stepwise through α-FeOOH (hexagonal close-packed structure) and γ-FeOOH (cubic close-packed structure). This chemical reaction involves the dehydration of

structural $H_2O$ and the partial reduction of $Fe^{3+}$ to $Fe^{2+}$. With high $H_2O$ volume (blue lines), the dehydration process is disfavored, leading to the assembly growth of a 2D lattice based on ligand incorporation.

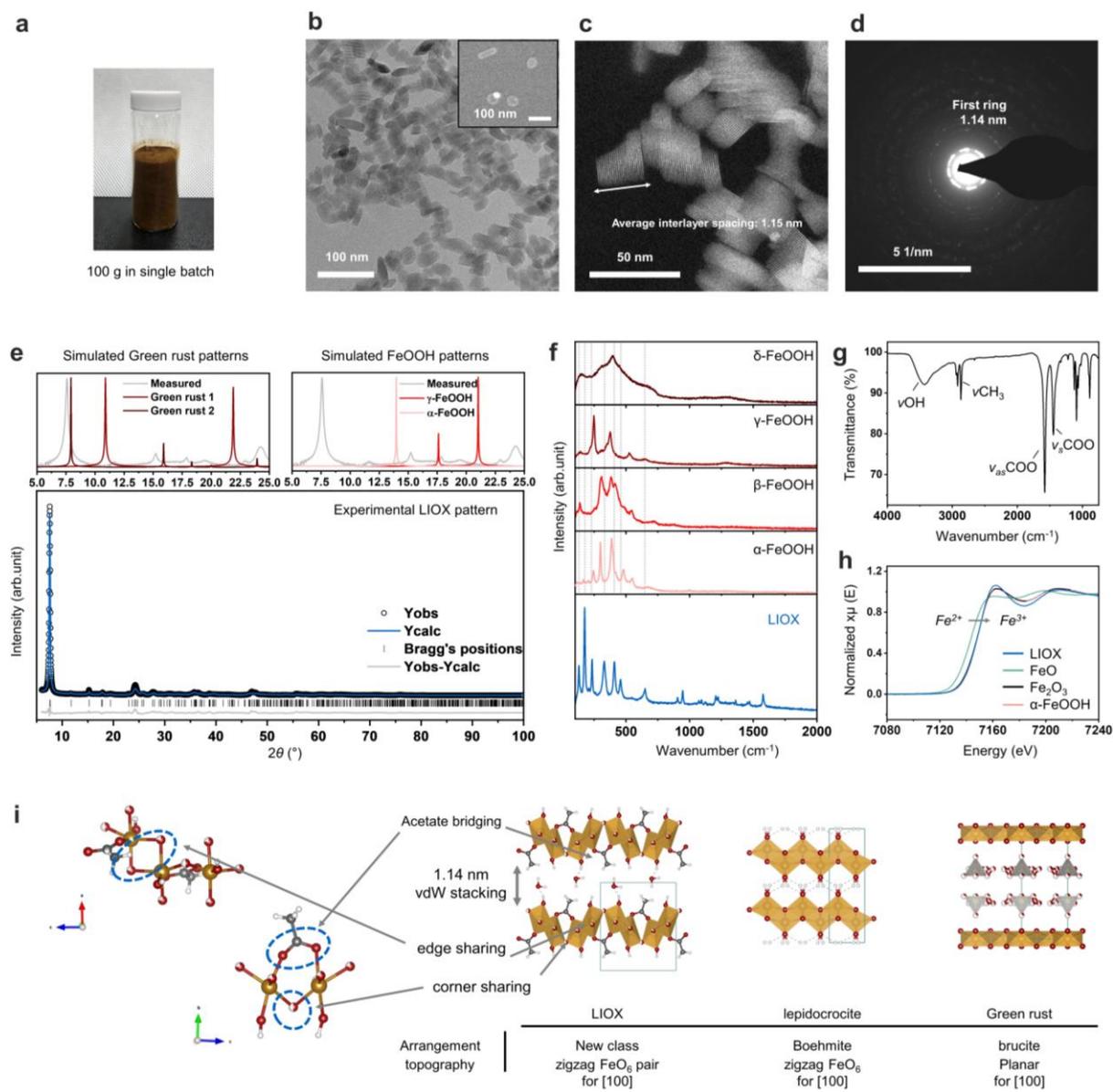

**Fig. 2. Atomic structure of LIOX nanoplatelets.** (**a**) Photograph of 100 g-scale samples synthesized in a single batch. (**b**) Bright–field TEM image. (**c**) Low-voltage STEM image measured at 80 kV. The intensity profile represent interlayer distances. See also Supplementary Fig. 5 and Supplementary Movie 1. (**d**) SAED pattern with intense diffraction ring at d = 1.14 nm. (**e**) Rietveld refinement of synchrotron HRXRPD measured with a wavelength of 1.5225 Å. The top panels compare the measured HRXRPD pattern of LIOX with simulated patterns of FeOOH polymorphs and *green rust* with a layered structure. Inset: (**f**) Raman spectra of LIOX and FeOOH polymorphs in the range below 2000 cm$^{-1}$. Grey dotted lines indicate the characteristic Raman shift of LIOX. (**g**) The FT-IR spectrum showing acetate coordination on the LIOX surface. (**h**) Fe K-edge XANES spectra of LIOX and other known iron compounds. (**i**) The structure of LIOX based on computational analysis of HRXRPD and DFT. Comparison of an atomic configuration of LIOX with known FeOOH phases of *lepidocrocite* and *green*

*rust 2* with a layered structure. Atoms are marked as dark orange for iron, red for oxygen, black for carbon, and white for hydrogen.

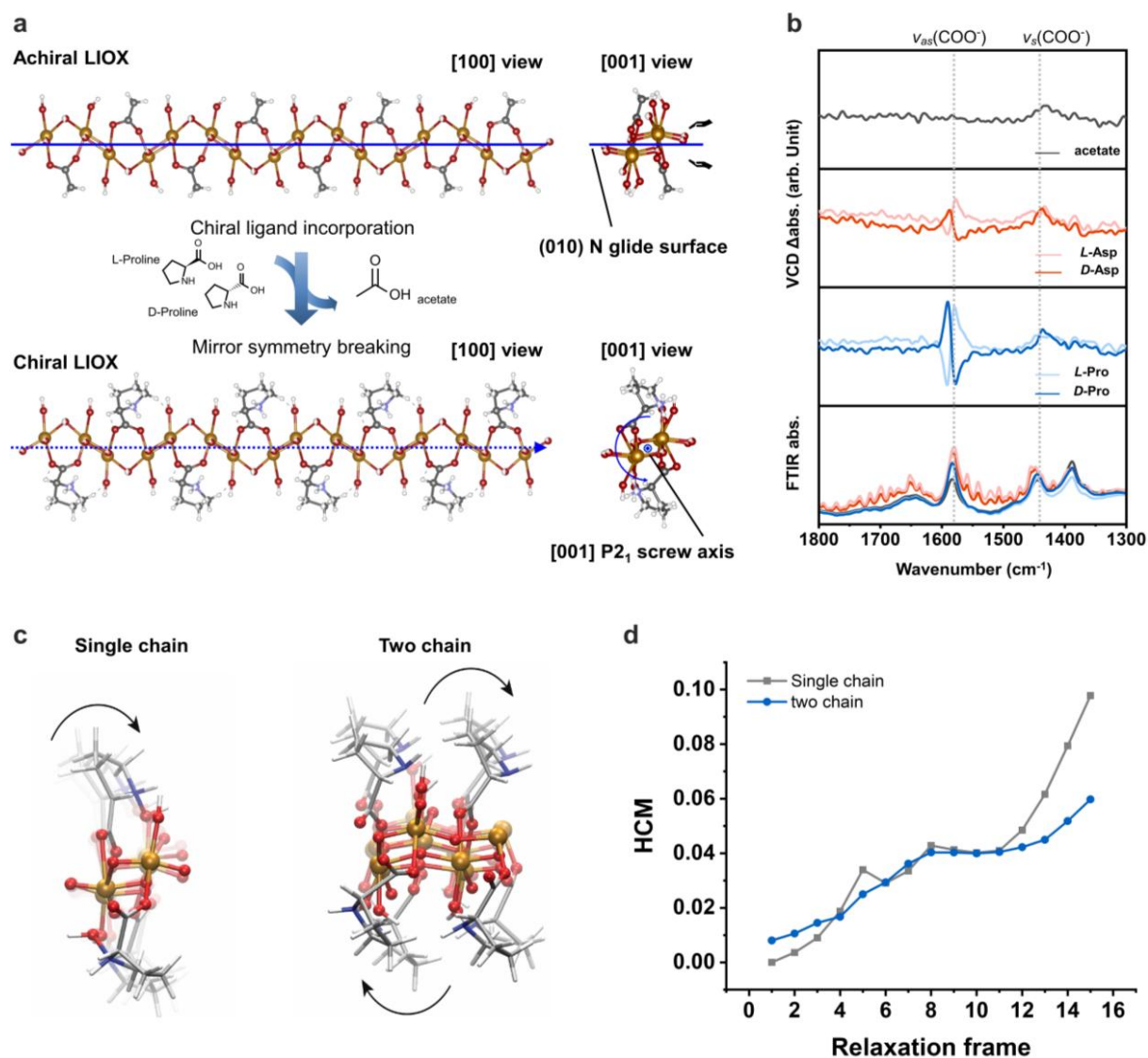

**Fig. 3. Ligand-induced chirality symmetry breaking.** (**a**) Atomic configuration model of achiral and chiral LIOX. Ligand exchange from acetate to proline during crystallization of LIOX induces mirror symmetry breaking and screw axis formation. (**b**) Infrared and VCD spectra of LIOXs intercalated with acetate (grey), *L*–Asp (light red), *D*–Asp (red), *L*–Pro (sky blue), and *D*–Pro (blue) LIOX. (**c**) MD simulation of helical distortion in FeOOH lattice of different zigzag unit numbers induced by chiral ligand attachment (see also Supplementary Movie 2). (**d**) HCM plot upon structural relaxation in MD simulation. It quantitatively demonstrates the gradual (albeit nonmonotonic) increase of chirality due to binding of chiral surface ligands to achiral LIOX crystal lattice.

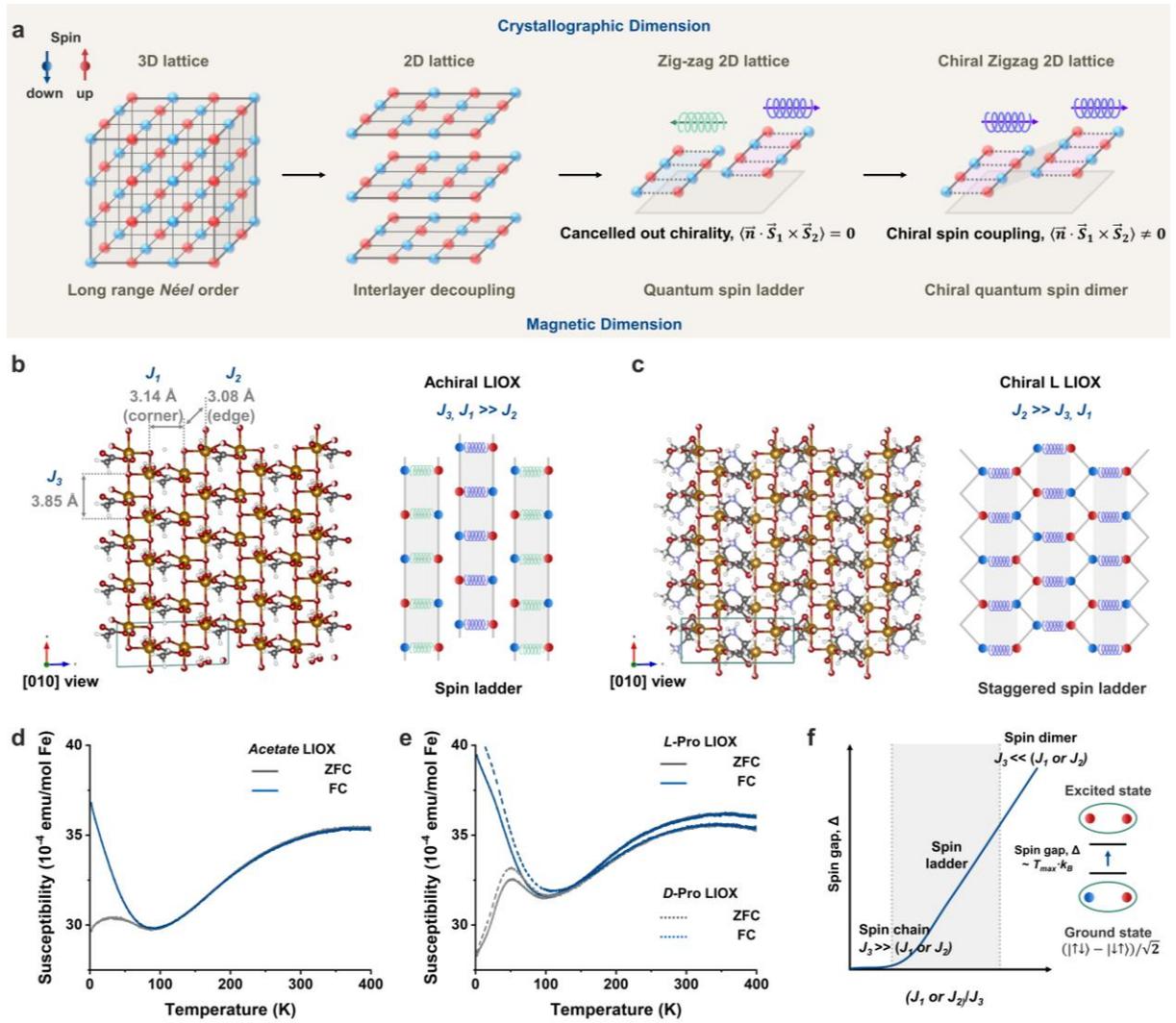

**Fig. 4. Quantum magnetism.** (**a**) Lattice engineering for 3D to quantum magnetic transition. LIOX is an ideal model for elucidating the topological and dimensional correlation between crystallography and quantum magnetism, as it allows for a series of lattice fractionalization steps, including: (i) 3D to 2D transition, (ii) zigzag topography, and (iii) chiral symmetry breaking. Blue and red circles indicate antiferromagnetically coupled spin dimers of the Fe sublattice. Purple (right-handed) and green (left-handed) helixes indicate the DMI $\vec{D} \cdot \vec{n}$, where *n*-glide symmetry flips the sign (cancels out the chirality) and two-fold screw rotation maintains the sign (chiral helicity) (**b** and **c**) Atomic configuration and topological spin structure of the 2D sheet (010) plane of achiral (b) and chiral (c) LIOXs. Schematic illustrations propose a spin ladder system without chiral coupling and a staggered spin ladder system with chiral coupling constructed based on DFT calculation. (**d** and **e**) Temperature-dependent magnetization (M–T) curves of achiral LIOX (d) and chiral LIOX (e) recorded by ZFC (black)/FC (blue) processes under external magnetic field (H) = 1000 Oe. (**f**) Spin-gapped state diagram depending on the spin coupling strength ratio between $J_1$, $J_2$, and $J_3$.

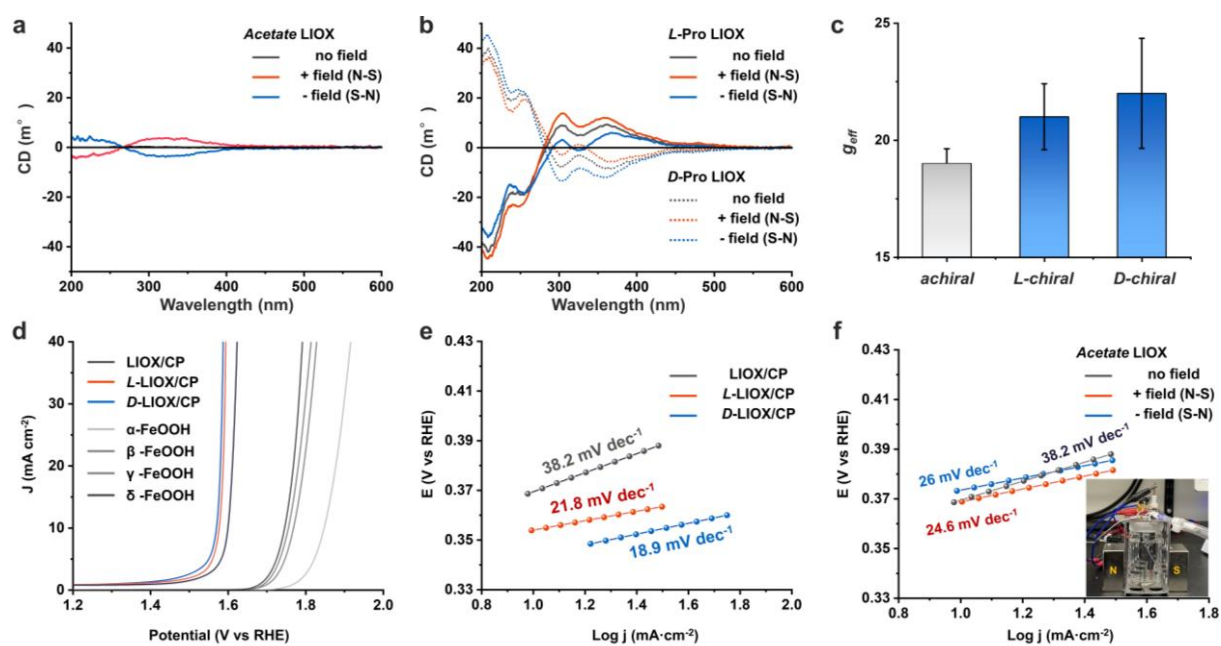

**Fig. 5. Spin-polarization effect of a colloidal quantum magnet on chiroptical and electrocatalytic properties.** (**a** and **b**) MCD spectra of acetate LIOX (a) and *L*– and *D*-Pro LIOXs (b). Applied magnetic fields are +1.6 T (N–S direction, red) and −1.6 T (S–N direction, sky blue). (**c**) Extracted $g_{eff}$ on spin-polarized photon absorption, indicating field-tuneability. (**d**) OER polarization curves of achiral LIOX, chiral LIOXs, and FeOOH polymorphs. (**e** and **f**) OER Tafel plots derived from the OER polarization curves of the corresponding catalysts; (e) chirality effect of achiral and chiral LIOXs with Pro surface ligands, and (f) magnetic field effect of achiral LIOX catalyst with acetate surface ligands.